\documentstyle[charm2000]{article}

\newcommand{\lsim}{\raisebox{-1mm}{$\stackrel{<}{\sim}$}}

\newcommand{\bx}[1]{\mbox{\boldmath $#1$}}
\catcode`\@=11
\def\eqalign#1{\null\,\vcenter{\openup\jot\m@th
\ialign{\strut\hfil$\displaystyle{##}$&$\displaystyle{{}##}$\hfil
     \crcr#1\crcr}}\,}
\catcode`\@=12

\begin{document}
\title{Hadrons with two heavy quarks}
\author{Jean-Marc Richard\\
{\sl Institut des Sciences Nucl\'eaires}\\
{\sl Universit\'e Joseph Fourier, CNRS-IN2P3}\\
{\sl 53, avenue des Martyrs, 38026 Grenoble Cedex, France}
}
\maketitle
\begin{abstract}
We review   the spectroscopy and some properties of hadrons
containing two charmed quarks, or more generally, two heavy quarks.
This includes heavy baryons such as $(bcu)$, and
possible exotic multiquark states.
\end{abstract}
\section{Introduction}
Baryons with two heavy quarks and one light quark, hereafter
denoted $(QQq)$, intimately
combine two extreme regimes of hadron structure. There is first the
slow relative motion of the two heavy quarks, very similar
to the quark--antiquark motion in charmonium and bottomonium.
In both cases, the heavy constituents experience an adiabatic
potential generated by the light degrees of freedom. The second
aspect of
 $(QQq)$ is the relativistic motion of the light quark $q$, which
is presumably very similar for $(ccq)$, $(bcq)$, and $(bbq)$,
providing
another example of heavy quark symmetry.

A rich spectrum  is expected. There are excitations
of the relative motion of the
two heavy  quarks in the lowest Born--Oppenheimer potential.
One can also  get excitations of the light quark, or a combined
excitation of
both
degrees of freedom.

The ground state of each flavour configuration cannot do anything but
decay weakly,
by disintegration of one of the heavy quarks,
and sometimes by  exchange of a $W$-boson between the constituents.
A variety of final states are accessible, with no, some, or more
Cabbibo
suppression. We have here an ideal laboratory for studying weak
interactions
and subsequent hadronisation.

If $(QQq)$ spectroscopy becomes accessible to experiment, it will
also
be possible to look at exotic mesons with two heavy quarks, $(QQ\bar
q\bar q)$.
They have been predicted to be stable on the basis of the flavour
independence
of the static interquark potential.
Other approaches have led to similar conclusions.
Current models gives stability for
ratios $(M/m)$ of quark masses corresponding to $(bb\bar q\bar q)$ or
higher.
However,
reasonable long-range forces might well push down this ratio, so that
some $(cc\bar q\bar q)$ could become serious candidates to stability.

In this review, I shall briefly summarize these aspects of
double heavy-flavour spectroscopy. General references are
\cite{FR1,FR2,JMRrep,Bagan} for $(QQq)$ spectroscopy
in potential models, \cite{Bagan,Savage} for
decays of these $(QQq)$, \cite{AdRiTa,Tetraq} for $(QQ\bar q \bar q)$
exotics in simple models, while a comparison with atomic physics is
attempted in
\cite{Hydrogen2}, and  another approach is discussed in
\cite{Tornqvist,ManoharQQqq}.
It is hoped that this Workshop will stimulate further investigations.
\section{Relations among ground state masses}
\label{se:ground}
The value of a peculiar $(QQq)$  mass is interesting only when
compared with
that of other flavour configurations. In the past, regularities have
been
noticed in the baryon spectrum, such as the Gell-Man--Okubo mass
formula, or
the equal-spacing rule of the decuplet. One possible interpretation
in the modern language is based on {\em flavour independence}. The
binding
potential is the same whatever quark experiences it. This property is
a
consequence of the gluons being coupled to the colour rather than
to the isospin, or
hypercharge, or mass of the quarks, at least before any relativistic
correction
is written down.
We shall come back on flavour independence in Sec.\ \ref{se:exotic},
and stress
the analogy with atomic physics, where the same $-1/r$ potential
binds
positronium, hydrogen and protonium atoms.

In the meson sector, we expect the lowest $(b\bar c)$ meson
approximately half between $J\!/\Psi$ and $\Upsilon$. In a
flavour-independent
potential, this is in fact a lower bound \cite{BeMa},
i.e., we have
\begin{equation}
\label{bc-cc-bb}
2(b\bar c)\ge (c\bar c)+ (b\bar b).
\end{equation}
If one knows the excitation spectrum of $(c\bar c)$ and $(b \bar b)$,
one can extract model-independent bounds on the average kinetic
energy
in the ground state, which governs the evolution of  the ground-state
energy
when
the reduced mass varies. This leads to a upper bound on
the lowest  $(b \bar c)$  state \cite{Bagan}, and
all predictions of realistic potentials nicely cluster near 6.26
GeV/$c^2$
\cite{Eichten-Quigg-bc}
in between the lower and the upper bounds provided by flavour
independence.

Similar regularity patterns are expected in the baryon sector
(the mathematics of the 3-body problem is of course more delicate
than
that of the 2-body one, and sometimes requires some mild conditions
on the shape of the confining potential, which are satisfied by all
current
models \cite{JMRrep}).  For instance, one expects an analogue of
(\ref{bc-cc-bb})
\begin{equation}
\label{cqq-ccq-qqq}
2(cqq)\ge (ccq)+(qqq)
\end{equation}
which leads to a upper bound $(ccq)\le 3.7\,$GeV for the c.o.m.\ of
the
ground-state multiplet of $(ccq)$. A upper bound can also be derived
for
$(ccs)$.
On the other hand, the convexity relation
\begin{equation}
\label{bcq-ccq-bbq}
2(bcq)\ge (ccq)+(bbq),
\end{equation}
cannot be tested immediately, as well as the even more
exotic-looking \cite{Martin-bcq}
\begin{equation}
\label{bcq-bbb-ccc-qqq}
3(bcq)\ge (bbb)+ (ccc) + (qqq),
\end{equation}
and its analogue with $q\rightarrow s$.
Of more immediate use is the relation
\begin{equation}
\label{bcq-bqq-cqq-qqq}
(bcq)\ge (bqq)+(cqq)-(qqq),
\end{equation}
which leads to a rough lower bound $(bcq)\ge6.9\,$GeV$/c^2$, if one
inputs the
following rounded and spin-averaged values: $(bqq)=5.6$, $(cqq)=2.4$,
and
$(qqq)=1.1\,$GeV$/c^2$.

To derive these inequalities, one  uses the Schr\"odinger equation,
even for
the
light quarks. Very likely, the regularities exhibited by
flavour-independent
potentials also hold in more rigourous QCD calculations and
in the experimental spectrum. Any failure of the above inequalities
would
be very intriguing.

Sometimes, one can be more precise, and derive inequalities that
include
spin--spin corrections, for instance relations between $J^P=(1/2)^+$
baryons with different flavour content. See \cite{JMRrep} for
details.

Another mathematical game triggered by potential models
consists of writing inequalities among meson and baryon masses.
The basic relation is \cite{JMRrep}
\begin{equation}
\label{meson-baryon}
2(q_1q_2q_3)\ge (q_1\bar q_2) + (q_2\bar q_3) +(q_3 \bar q_1),
\end{equation}
obtained by assuming that the potential energy operators
fulfill the following inequality
\begin{equation}
\label{pot-mes-bar}
2V_{qqq}({\bf r}_1,{\bf r}_2,{\bf r}_3)\ge \sum_{i<j}V_{q\bar
q}(|{\bf
r}_i-{\bf r}_j|),
\end{equation}
which holds (with equality) for a colour-octet exchange, in
particular
one-gluon exchange, and for the simple  model
\begin{equation}
\label{string}
V_{q\bar q}(r)=\lambda r,\qquad V_{qqq}=\lambda \min_J(d_1+d_2+d_3)
\end{equation}
where $d_i$ is the distance from the $i$-th quark to a junction $J$
whose
location is adjusted to minimize $V_{qqq}$ \cite{Y-shape}.
We obtain for instance
\cite{FR1} $(ccq)\ge 3.45\,$ GeV$/c^2$ for the $(1/2)^+$ state.
This is rather crude, not surprisingly. Years ago, Hall and Post
\cite{Post} pointed out in a different context
that the pairs are not at rest in a 3-body bound state, and that
their
collective
kinetic energy is neglected in inequalities of type
(\ref{meson-baryon}).
\section{Spectrum of doubly flavoured baryons}
\label{se:excitation}
Computing the $(QQq)$ energies in a given potential model does not
raise
any particular difficulty. The 3-body problem is routinely solved by
means of
the
 Faddeev equations or variational methods.
On the other hand,  successful approximations  often shed some
light on the dynamics. In particular, the
Born--Oppenheimer method works very well
for large  ratios $(M/m)$ of the quark masses. At
fixed $QQ$ separation $R$, one solves the 2-centre
problem for the light quark $q$. The energy of $q$ is added to the
direct $QQ$
interaction to generate the effective potential $V_{QQ}(R)$ in which
the
heavy quarks evolve. One then computes the $QQ$
 energy and wave function. Note that one can remove
the centre-of-mass motion exactly, and also estimate the hyperfine
corrections.

The physics behind the Born--Oppenheimer approximation is rather
simple.
 As the heavy quarks move slowly, the light degrees of freedom
readjust
themselves to their lowest configuration (or stay in the same $n$-th
excitation,
 more generally).
At this point, there is no basic difference with quarkonium.
The $Q\,\overline{\!Q}$
potential does not represents an elementary process. It can be viewed
as
the effective
interaction generated by the gluon field being in its ground-state,
for a given
$Q\,\overline{\!Q}$ separation.

The results shown in Table \ref{Table1} come from the simple
potential
\begin{equation}
\label{eq:potential}
V={1\over2}\sum_{i<j}\left[A+Br_{ij}^\beta+{C\over
m_im_j}\bx{\sigma}_i\cdot
\bx{\sigma}_j\delta^{(3)}({\bf r}_{ij})\right],
\end{equation}
with parameters $\beta=0.1$, $A=-8.337$, $B=6.9923$, $C=2.572$,
in units of appropriate powers of GeV.
The quark masses are $m_q=0.300$, $m_s=0.600$, $m_c=1.905$ and
$m_b=5.290\,$GeV.
The $1/2$ factor is a pure convention, although reminiscent from the
discussion
of inequalities (\ref{meson-baryon}) and (\ref{pot-mes-bar}).
The smooth central term
can  be seen as a handy interpolation between the short-range
Coulomb regime modified by asymptotic-freedom corrections
and an elusive linear regime screened by pair-creation effects.
The spin-spin term is treated at first order to estimate $M_0$.
This model fits all known gound-state baryons with at most one heavy
quarks.
\begin{table}[h]
\caption{\label{Table1} Masses, in GeV,  of
\protect{$(QQq)$} baryons in a simple potential
model. We show the spin-averaged mass \protect{$\,\overline{\!M}$},
and the mass \protect{$M_0$} of the lowest state with
\protect{$J^P=(1/2)^+$}.}
\begin{center}
\begin{tabular}{ccccccc}
State&$ccq$&$ccs$&$bcq$&$bcs$&$bbq$&$bbs$\\
$\,\overline{\!M}$&3.70&3.80&6.99&7.07&10.24&10.30\\
$M_0$&3.63&3.72&6.93&7.00&10.21&10.27\\
\end{tabular}
\end{center}
\end{table}

A more conventional Coulomb-plus-linear potential was used
in Ref.\ \cite{FR1}, with similar results. One remains, however,  far
from
the large number of models available for $(b\bar c)$
\cite{Eichten-Quigg-bc},
and the non-relativistic treatment of the light quark might induce
systematic
errors. The uncertainty is then conservatively estimated to be
$\pm50\,$MeV,
as compared to $\pm20\,$MeV for $(b\bar c)$.
Note also that the $b$-quark mass $m_b$ is tuned
to reproduce the experimental mass of $\Lambda_b$ at 5.290 GeV$/c^2$,
and this latter value is not firmly established.

The Born--Oppenheimer framework leaves room for improvements.
 A relativistic treatment of the light quark was attempted in
\cite{FR1},
using the bag model. For any given $QQ$ separation, a bag is
constructed in
which the light quark moves. The shape of the bag is adjusted to
minimize the energy. In practice, a spherical approximation is used,
so that the radius is the only varying quantity. The energy of the
bag and
light quark  is interpreted as the effective $QQ$ potential.
Unlike the rigid MIT
cavity, we have a self-adjusting bag, which follows the $QQ$ motion.
Again, this is very similar to the
bag model picture of charmonium \cite{Kuti}.

Unfortunately, there are variants in the bag model, with different
values
of the parameters, and with or without corrections for
the centre-of-mass motion. These variants lead to rather different
values
for the $(ccq)$ masses \cite{FR1}.
This contrasts with the clustered shoots of potentials models,
and deprives the bag model
of predictive power in this sector of hadron spectroscopy.

It is hoped that the $QQ$ potential will be calculated by
lattice or sum-rule methods.

The excitation spectrum of $(QQq)$ baryons has never been calculated
in great
detail,
at least to our knowledge. In Ref.\ \cite{FR1}, an estimate is
provided for the
spin
excitation (ground state with $J^P=(3/2)^+$), the lowest
negative-parity level,
and
the  radial  excitation of the ground state.

The spin excitation is typically 100 MeV above the ground state, and
thus
should
decay radiatively, with a $M1$ transition. The orbital and radial
excitations
of
$ccq)$ are unstable, since they can emit a pion. The radial
excitation of
$(ccs)$ can
decay into $(ccq)+K$, but the orbital excitation cannot, and thus
should be
rather
narrow, since restricted to $(ccs)+\gamma$, or to the
isospin-violating
$(ccs)+\pi^0$.
\section{Decay of heavy baryons}
\label{se:decay}
The ground state of $(QQq)$ decays weakly, with a great variety of
final
states.
 For instance, the remaining
heavy flavour can stay in the
baryon, or join the meson
sector. Moreover, we have
Cabibbo allowed, suppressed, or
doubly suppressed modes. We
refer to Savage et al.\ \cite{Savage} for a
comprehensive survey of 2-body channels
of interest.

Inclusive decay rates are also of great importance.
The difference between the $D^0$ and $D^+$ lifetimes tells us that
the charmed quark, while decaying, does not ignore its environment.
The
main process is $c\rightarrow s+W$, and $W\rightarrow u\bar d$ for
hadronic modes, but one should also consider $W$-exchange
contribution for
$D^0$, interferences between the
two $\bar d$ in $D^+$ decay, $c\bar s$ annihilation for $D_s$, etc.

The lifetimes of single-charm baryons have been analysed by Guberina
et al.\
\cite{Ruckl}. The annihilation diagram requires antiquarks from the
sea, and
presumably does not play a very important role. On the other hand,
$W$-exchange
does not suffers
from helicity suppression.  We  also have two types of interferences:
between constituent
$u$ and $u$ from $W$ decay, and between constituent $s$ and $s$ from
$c$
transmutation.
The predictions of \cite{Ruckl}
\begin{equation}
\label{single-c-lifetime}
\tau\left(\Omega_c^0\right)\lsim\tau\left(\Xi_c^0\right)
<\tau\left(\Lambda_c^+\right)<\tau\left(\Xi^+_c\right),
\end{equation}
seems confirmed by recent data. If one
extrapolates their analysis toward the
$(ccq)$ sector, one predicts \cite{FR1}
\begin{equation}
\label{double-c-lifetime}
\tau\left(\Xi_{cc}^+\right)
<\tau\left(\Omega_{cc}^+\right)<\tau\left(\Xi^{++}_{cc}\right).
\end{equation}

\section{Exotic mesons with two heavy quarks }
\label{se:exotic}
The situation and the perspectives for the pentaquark will be
reviewed by
Moinester \cite{Moinester}. The pentaquark is an exotic baryon
($B=1$) with
charm (or heavy flavour) $C=-1$, i.e., a $(\,\overline{\!Q}qqqq)$
struture.
We shall discuss another possible multiquark, the tetraquark, with
$B=0$ and
$C=2$.
The main difference, besides these quantum numbers, is that the
pentaquark is
tentatively bound by chromomagnetic forces, while the tetraquark uses
a
combination
 of flavour-independent chromoelectric forces, and Yukawa-type of
long range
forces.

Recently, T{\"o}rnqvist \cite{Tornqvist}, and Manohar and Wise
\cite{ManoharQQqq}
studied pion-exchange between heavy mesons, and stressed that, among
others,
some
$DD^\star$ and $BB^\star$ configurations experience attractive
long-range
forces. By
itself, this Yukawa potential seems unlikely to bind $DD^\star$, but
might
succeed
for the heavier $BB^\star$ system.

Years ago, Ader et al.\ \cite{AdRiTa} showed that $(QQ\bar q\bar q)$
should
become
stable for very large quark-mass ration $(M/m)$, a consequence of the
flavour independence of chromoelectric forces. The conclusion was
confirmed in
subsequent studies \cite{Tetraq}.

In the limit of large $(M/m)$,  $(QQ\bar q\bar q)$ bound states
exhibit a
simple
structure. There is a localized $QQ$ diquark with colour $\bar 3$,
and this
diquark
forms a colour singlet together with the two $\bar q$, as in every
flavoured
antibaryon. In other words, this multiquark uses well-experienced
colour
coupling,
unlike speculative mock-baryonia or other states proposed in ``colour
chemistry''
\cite{Chan}, which contain  clusters with colour 6 or 8.

The stability of $(QQ\bar q\bar q)$ in flavour-independent potentials
is
analogous to
that of the hydrogen molecule \cite{Hydrogen2}.
If one measures the binding in units of the threshold energy, i.e.,
the energy
of two
atoms, one notices that the positronium molecule $(e^+e^+e^-e^-)$
with equal
masses
is bound by only 3\%, while  the very asymmetric hydrogen reaches
17\%. This
can be understood by writting
the molecular Hamiltonian as
\begin{equation}
\label{eq:hydrogen}
\eqalign{
H=&H_{\rm S}+H_{\rm A}\cr
=&\left({1\over4M}+{1\over4m}\right)\left({\bf p}_1^2+{\bf p}_2^2
+{\bf p}_3^2+{\bf p}_4^2\right)+V\cr
+&\left({1\over4M}-{1\over4m}\right)\left({\bf p}_1^2+{\bf p}_2^2
-{\bf
p}_3^2-{\bf
p}_4^2\right) }
\end{equation}
The Hamiltonian $H_{\rm S}$, which is symmetric under charge
conjugaison, has
the
same threshold as $H$, since only the inverse reduced mass
$(M^{-1}+m^{-1})$
enters the
energy of the $(M^+m^-)$ atoms. Since $H_{\rm S}$ is nothing but a
rescaled
version
of the Hamiltonian of the positronium molecule, it gives 3\% binding
below the
threshold. Then the antisymmetric part $H_{\rm A}$ lowers the
ground-state
energy of $H$, a
simple consequence of the variational principle.

In simple quark models without spin forces, we have a similar
situation. The
equal
mass case is found unbound, and $(QQ\bar q\bar q)$ becomes stable,
and more and
more
stable, as $(M/m)$ increases.
One typically  needs $(bb\bar q\bar q)$, with $q=u$ or $d$, to
achieves binding
with the
nice diquark clustering we mentioned. However, if one combines this
quark attraction with the long-range Yukawa forces, one presumably
gets binding
for
$(bb\bar q\bar q)$ with $DD^\star$ quantum numbers.
A more detailed study is presently under way \cite{Tornqvist2}.

The experimental signature of tetraquark heavily depends on its exact
mass.
Above $DD^\star$, we have a resonance, seen as a peak in the
$DD^\star$ mass
spectrum.
Below $DD^\star$, one should look at $DD\gamma$ decay of tetraquark.
If it lies
below
$DD$, then it is stable, and decays via weak interactions, with a
lifetime
comparable
to that of other charmed particles.

\section*{Acknowledgements}
I would like to thank C. Quigg, S. Narison and D. Kaplan for having
revived my interest into double-flavour spectroscopy,
the organizers of the ``Charm2000'' Workshop for their invitation,
and T. Mizutani for several discussions and a visit at Virginia
Tech., where this review was prepared.
%
%

\end{document}